\documentclass[conference,10pt]{IEEEtran}
\usepackage[T1]{fontenc}
\usepackage[utf8]{inputenc}
\usepackage[english]{babel}
\usepackage{amsmath}
\usepackage{amsfonts}
\usepackage{amssymb}
\usepackage[nolist]{acronym}
\usepackage{graphicx}
\usepackage{psfrag}
\usepackage{epstopdf}
\usepackage{authblk}

\title{Design of MLSD-Based Receivers \\ for Short-Range Optical Communications \\ Using the Volterra Expansion}

\author{Raquel G. Machado, Beatrice Tomasi, Hartmut Hafermann and Stefano Tomasin \\
Mathematical and Algorithmic Sciences Lab, Huawei Technologies Co. Ltd., France. \\
 E-mail: \texttt{\{raquel.machado, beatrice.tomasi, hartmut.hafermann\}@huawei.com} \\
 \texttt{tomasin@dei.unipd.it}}

\begin{document}

\maketitle

\renewcommand{\IEEEiedlistdecl}{\IEEEsetlabelwidth{S
ONET}}
\begin{acronym}

\acro{IM/DD} {Intensity Modulation/Direct Detection}
\acro{WDM} {Wavelength Division Multiplexing}
\acro{AWGN}{Additive White Gaussian Noise}
\acro{ASE}{Amplified Spontaneous Emission}
\acro{BER}{Bit Error Rate}
\acro{CD}{Chromatic Dispersion}
\acro{PMD}{Polarization-Mode Dispersion}
\acro{ODSP}{Optical Digital Signal Processing}
\acro{NRZ}{Non-Return-to-Zero}
\acro{OOK}{On-Off Keying}
\acro{MLSE}{Maximum Likelihood Sequence Estimator}
\acro{MLSD}{Maximum Likelihood Sequence Detector}
\acro{DMT}{Discrete Multi-Tone}
\acro{MF}{Matched Filter}
\acro{WF}{Whitening Filter}
\acro{DDFSE}{Delayed Decision Feedback Sequence Estimator}
\acro{ISI}{Inter-Symbol Interference}
\acro{OSNR}{Optical Signal-to-Noise Ratio}
\acro{pdf}{Probability distribution function}
\acro{VA}{Viterbi Algorithm}
\acro{VD}{Viterbi Decoder}
\acro{ST-WMF-MLSD}{Space-Time Whitened Matched Filter MLSD}
\acro{ADC}{Analog-to-Digital Converter}
\acro{OOK}{On-Off Keying}
\acro{PAM}{Pulse Amplitude Modulation}

\end{acronym}
\renewcommand{\IEEEiedlistdecl}{\relax}

\begin{abstract}
Maximum Likelihood Sequence Detectors (MLSD) have been largely used to mitigate the \ac{CD} in \ac{IM/DD} optical communication systems. For practical applications, the high complexity of the receivers remains an important issue. In this paper, we analyze the design of MLSD-based receivers using the Orthogonal Volterra Kernel Model for IM/DD optical communication systems in Metro Optical applications. We discuss the impact in complexity and performance of the main parameters of the model and provide three design options for the MLSD-based receiver. Finally we provide numerical simulations showing the \ac{BER} performances of the three considered designs for both \ac{OOK} and higher order \ac{PAM}. 
\end{abstract}


\section{Introduction}

High capacity and low cost are the main design goals in short reach optical communication systems. \ac{IM/DD} technology provides a popular solution \cite{cartledge2014100}. The non-linearity introduced by the direct detection and its interaction with \ac{CD}, however, is a major issue in the design of \ac{IM/DD} optical communication systems.
It is therefore necessary to design low-complexity receivers able to deal with these impairments. 

Maximum likelihood sequence detection has been increasingly used to mitigate the \ac{CD} and \ac{PMD} generated by the optical fiber \cite{MLSD_optical1,MLSD_optical2,channelest}. Specifically, the \ac{CD} phenomenon can be modeled as \ac{ISI} \cite{savory2005robust} and a \ac{MLSD} receiver with a \ac{VD} was originally developed for equalizing inter-symbol interference (ISI) based on a linear channel model under additive white Gaussian noise (AWGN) \cite{cherubini2003algorithms}. Given the relatively short distances in Metro Optical networks, the memory introduced through \ac{CD} is sufficiently small to justify the usage of \ac{MLSD}-based receivers \cite{practical}.

Recently, \cite{maggio2014reduced, maggio2015maximum} used the Volterra-series Expansion theory \cite{volterra} to design a reduced-complexity MLSD receiver for optical channels, the \ac{ST-WMF-MLSD}. The proposed receiver structure leverages the energy compression provided by an Orthogonal Volterra Kernel model to decrease the memory required by the \ac{VD} at the receiver. However, even though the complexity gains regarding the memory in the \ac{VD} are significant for fibers with length of several hundreds kilometers, the relative gains decrease for shorter ranges, between $30-100$km. Within this context, the impact in complexity of the filter-bank at the input of the decoder and the metric calculations become relevant in comparison to the complexity generated by the memory of the decoders.

In this paper, we analyze the design of MLSD-based receivers using the Orthogonal Volterra Kernel Model for IM/DD optical communication systems in short-range Metro Optical applications. We discuss the impact in complexity and performance of an MLSD-based receiver regarding important parameters of the Orthogonal Volterra Kernel system model and provide three design options for the receiver structure. Finally, we provide numerical simulations showing the BER performances of the considered designs for both OOK and higher-order PAM modulation, motivating the usage of the proposed receivers in a higher data rate scenario without requiring a significant increase in complexity.

\section{Optical Channel Model}

\begin{figure}[t!]
  \centering
  \psfrag{B}{$\cal{A}$}
  \psfrag{C}{$|\centerdot|^2$}   
    \includegraphics[width=0.5\textwidth]{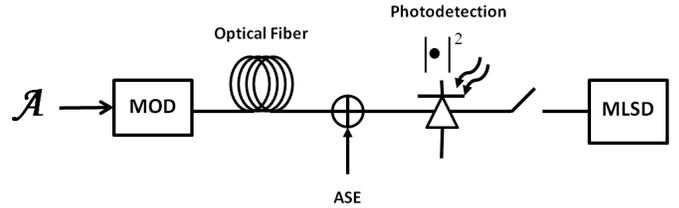}
 \caption{Optical communications system model with a \ac{MLSD}-based receiver.}
\label{fig:phy_sys}
\end{figure}

In Fig. \ref{fig:phy_sys} the model of the considered \ac{IM/DD} system is provided. Baseband digital symbols $\mathcal{A}$ modulate the light source at a symbol rate of $1/T$. The modulated signal is then propagated through a linear optical fiber characterized by the frequency response

\begin{equation} \label{eq:fiber}
O(\omega)= \exp{ \left( -j \frac{\lambda^2DL}{4\pi C}\omega^2  \right)},
\end{equation}
where $\lambda$ is the wavelength of the optical carrier, $C$ is the speed of light and DL is the \ac{CD} of the fiber. To compensate for the attenuation of the optical signal, optical amplifiers are deployed along the fiber, introducing \ac{ASE} noise, here modeled as \ac{AWGN} in the optical domain. The received optical signal is then transformed into electrical current which is proportional to the power of the  optical field. In this case, we model the photodetection process as a memoryless modulus-square operation.

According to the model and \cite{opticalISI}, we are able to model the post-detection analog signal $y(t)$ in terms of the transmitted symbols as
\begin{equation}
\left| \sum_k a_k g(t-kT) \right|^2, \label{digipower}
\end{equation}
where $a_k$ is the $kth$ real symbol at the input of the nonlinear channel and $g(t)$ is the optical pulse propagated through the optical channel $o(t)$ in the continuous domain. Taking this model into consideration, we can expand \eqref{digipower} using the modulus-square identity:
\begin{align}\label{eq:nonlinear}
y(t) &= \left(\sum_k a_k g(t-kT) \right)\left(\sum_{\ell} a_{\ell} g(t-\ell T) \right)^{*} \nonumber \\
&= \sum_{k} a_k^2 |g(t-kT)|^2  \\ 
& \quad + \sum_{\ell \neq k} \sum a_{\ell}  a_k g^{*}(t-kT)g(t-\ell T), 
\end{align}
where $(\cdot)^{\ast}$ denotes complex conjugate. After rearranging the terms and making the change of variables: $\ell = k+m$, it is possible to re-write \eqref{eq:nonlinear} as:
\begin{equation} \label{eq:simple_volt}
y(t) = \sum_{k} a_k^2 f_0 + \sum_{m>0} \sum_{k} a_k a_{k+m} f_{m}(t- kT)
\end{equation}
where
\begin{align*}
f_0 &= |g(t)|^2 \\
f_{m}(t)&=2Re\{ g(t)g^{*}(t-mT) \}.
\end{align*}

Since the dominant non-linearity present in the \ac{IM/DD} system considered in this paper comes from the photodection process, here modeled by the modulus-square operation,  \eqref{digipower} can be exactly expanded to \eqref{eq:simple_volt}. Also, the relationship of the symbol-pairs and the defined $f_m$ functions mirror the definition of a second-order Volterra expansion. In this sense, examining \eqref{eq:simple_volt} through the Volterra expansion framework, the terms can be defined as second-order Volterra Kernels, being $f_0$ the linear kernel and $f_m$ the kernels related to nonlinear interactions between symbols $m$ periods apart. Note that the kernels $f_0$, $f_1$, ..., $f_m$ are in general not orthogonal.

\section{Optimal MLSD receiver}
Even though the second-order Volterra kernel representation of the considered optical channel given by \eqref{eq:simple_volt} is exact, other ways of modeling the non-linear signal $y(t)$ may become more advantageous considering specific features of a \ac{MLSD}-based receiver. According to \cite{maggio2014reduced}, a viable alternative representation of the optical channel can be achieved by using orthogonal kernels. In this new representation, the functions that represent the Volterra kernels are submitted to an orthogonalization process that guarantee that most of the energy of the signal is present in fewer orthogonal kernels, eliminating redundancies, and generating a space-compression phenomenon.

\begin{figure}[t!]
  \centering
    \includegraphics[width=0.5\textwidth]{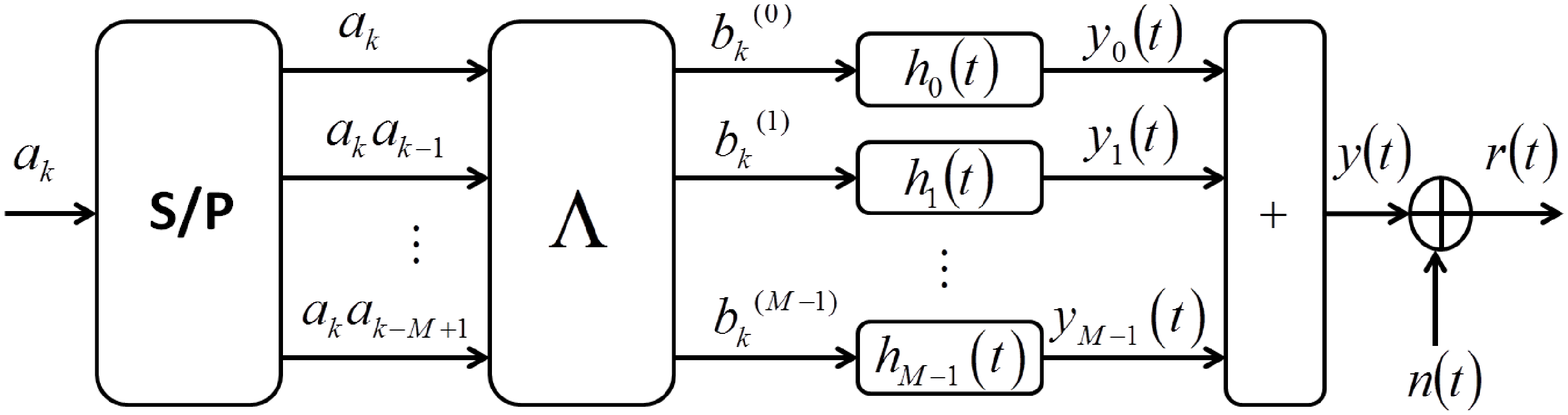}
 \caption{Transmission model using Volterra Kernels expansion.}
 \label{fig:volt_trans}
\end{figure}

The orthogonalization process presented in \cite{maggio2014reduced} begins by choosing the first pivoting kernel $h_0[i]=h_0[t]|_{t=iT_s}$, from the $f_m$. The orthogonalization, which is similar to a Gram-Schmidt process, has $M$ steps, where $M$ is equal to the number of kernels used to model \eqref{eq:simple_volt}. At each step, a pivoting kernel is chosen and its projection onto all the remaining kernels. From the projection theorem, the other kernels can be expressed in terms of the projection of the pivoting kernel onto them and the respective projection error. For example, in the first step of the orthogonalization process we have:
\begin{equation} \label{eq:projection}
h_m[i] = f_m[i] - \sum_{n}^{N} \lambda_n^{(0,m)}h_0[i-nR], 
\end{equation}
where the summation term is the projection of $f_m[i]$ onto the pivoting kernel $h_0[i]$. In matrix format, we can write \eqref{eq:projection} as:
\begin{equation} \label{eq:proj_mat}
\mathbf{h}_m =  \mathbf{f}_m  - \mathbf{H}_0 \boldsymbol{\lambda}_{(0,m)},
\end{equation}
where
\begin{align*} \small
&\mathbf{H}_0=\\
&\begin{bmatrix}
h_0[i_0] & h_0[i_0-R] & \hdots & h_0[i_0-(N-1)R] \\
h_0[i_1] & h_0[i_1-R] & \hdots & h_0[i_1-(N-1)R] \\
\vdots & \ddots & \hdots & \vdots \\
h_0[i_{L-1}] & h_0[i_{L-1}-R] & \hdots & h_0[i_{L-1}-(N-1)R]
\end{bmatrix}
\end{align*}
\begin{equation*}
\boldsymbol{\lambda}_{(0,m)} = \begin{bmatrix}
\lambda_{0}^{(0,m)} &
\lambda_{1}^{(0,m)} &
\dots &
\lambda_{N-1}^{(0,m)}
\end{bmatrix}^T .
\end{equation*}
With \eqref{eq:proj_mat}, we calculate $\boldsymbol{\lambda}_{(0,m)}$, based on the projection of $f_m[i]$ on $h_0[i]$, with the pseudo-inverse of $\mathbf{H}_0$:
\begin{equation*}
\boldsymbol{\lambda}_{(0,m)}= (\mathbf{H}_0^{H}\mathbf{H}_0)^{-1}\mathbf{H}_0^{H}\mathbf{f}_m
\end{equation*}
where $(\cdot)^{H}$ means Hermitian. The projection coefficients $\boldsymbol{\lambda}_{(*,m)}$ determine the relationship between the pivoting kernel and the other kernel functions guaranteeing their orthogonality. Fig. \ref{fig:volt_trans} illustrates the model of the transmitted signal $y(t)$ using the calculated mutually orthogonal kernels $h_0(t)$, $h_1(t)$, ..., $h_{M-1}(t)$.

\begin{figure}[t!]
  \centering
    \includegraphics[width=0.5\textwidth]{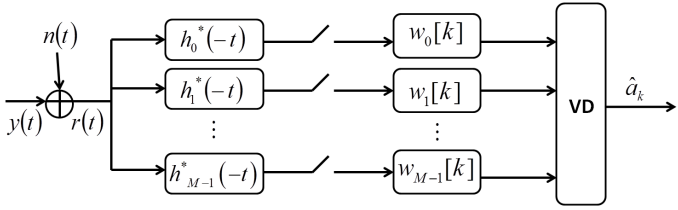}
 \caption{Optimal MLSD structure using the Orthogonal Volterra Kernel model for the transmitted signal $y(t)$.}
\label{fig:st_wmf_mlsd}
\end{figure}

Given the orthogonality property of the calculated kernels, and the fact that the optimal \ac{MLSD}-based receiver implemented with a \ac{VD} comprises a \ac{MF} and a \ac{WF} \cite{cherubini2003algorithms}, it is possible to design a receiver structure to implement the optimal MLSD receiver from the orthogonal Volterra Kernel transmission model. Fig. \ref{fig:st_wmf_mlsd} shows the optimal MLSD receiver implementation based on the orthogonal Volterra kernel model for the transmitted signal. In addition to the filter-bank structure, it is also necessary to use multidimensional Euclidean branch metrics in the \ac{VD} to implement the optimal MLSD receiver. In \cite{maggio2014reduced}, this structure is called \ac{ST-WMF-MLSD}.

\section{Orthogonal Volterra Kernel Model Parameters}
\label{sec:parameters}
Ideally, the number of kernels expanded from \eqref{eq:simple_volt} is infinite. In reality, the number of $2^{nd}$-order kernels considered for the expansion is finite, and we will define it as $M$:
\begin{equation} \label{eq:2kvolt_model}
y^{V}(t) = \sum_{k} |a_k|^2 f_{0}(t- kT) + \sum_{k} \sum_{m=1}^{M-1} a_k a_{k+m} f_{m}(t- kT).
\end{equation}
where $y^{V}(t)$ represents the signal modeled according the $2^{nd}$-order Volterra Kernels model.
In this case, $M$ will determine the \emph{modeling mismatch} between the modulus square signal presented in \eqref{digipower} and the $2^{nd}$-order Volterra Kernel expansion in \eqref{eq:2kvolt_model}.
After the orthogonalization process, \eqref{eq:2kvolt_model} can be written as:
\begin{align} \label{eq:orth_model}
y^{O}(t) &= \sum_{m=0}^{M-1} \sum_{k} b_k^{m}h_{m}(t- kT) \\
&= \sum_{m=0}^{M-1} \sum_{k} \Lambda(a_k,a_ka_{k+1},...,a_ka_{k+M-1} ) h_{m}(t- kT), \nonumber 
\end{align}
where $y^{O}(t)$ represents the signal modeled according the Orthogonal Volterra Kernel model, $\Lambda$ is the mapping that describes the relationship between the pairs of symbols $a_k,a_ka_{k+1},...,a_ka_{k+M-1}$ and the complex symbols $b_k^{m}$, which guarantees the orthogonality of the kernels $h_m(t)$. 

However, unlike the model in \eqref{eq:2kvolt_model}, the kernels in \eqref{eq:orth_model} are excited by different symbol pairs, depending on the order of the pivoting kernels chosen at each orthogonalization step. For example, if we choose $M=3$, \textit{i.e.}, we use a $3$-kernel expansion, and the pivoting kernels are chosen in the order $\{2,1,0\}$, the orthogonal kernel expansion is:
\begin{align*} 
y^{O}(t) &= \sum_{k} \Lambda(a_k,a_ka_{k+1},a_ka_{k+2} ) h_{0}(t- kT) \\
&+ \sum_{k} \Lambda(a_k,a_ka_{k+1}) h_{1}(t- kT) + 
\sum_{k} \Lambda(a_k) h_{2}(t- kT).
\end{align*}

Note that the first kernel is excited by the mapping of all of the three considered symbol-pairs, the second kernel is excited by two symbol-pairs and the third kernel by only one. This mapping shifts most of the energy present in the symbols to the first term of the expansion. In this sense, the modeling mismatch of discarding the kernels $h_1(t)$ and $h_2(t)$ for an eventual implementation is less significant compared to when only 1 kernel is considered in \eqref{eq:2kvolt_model}. This feature of the orthogonal Volterra expansion provides extra flexibility when choosing kernels in the model to implement the receiver, especially if complexity in the receiver is an issue.

Thus, when implementing the optimal receiver shown in Fig.~\ref{fig:st_wmf_mlsd}, it is important to correctly choose $M$ to decrease the modeling mismatch. In addition, it is possible to also choose the number of branches considered in the filter bank in order to decrease the receiver complexity, by truncating the system models represented by \eqref{eq:2kvolt_model} and \eqref{eq:orth_model}. In the rest of the paper, we denominate the total number of branches considered in the receiver as $U$.  

We define $y^{O}_{(M,U)}(t)$ as the signal modeled according the Orthogonal Volterra Kernel model obtained by orthogonalization of $M$ kernels and subsequent truncation to $U$ kernels. In analogy, we define $y^{V}_{(M,U)}(t)$ as the signal modeled according the $2^{nd}$-order Volterra Kernel model. Note that here $M$ is redundant, since the result depends only on the final number $U$ of terms kept in the expansion.

To evaluate the modeling mismatch between the modulus-square signal and $y^{V}_{(M,U)}(t)$ and $y^{O}_{(M,U)}(t)$, we use a figure of merit that we call signal to mean square error ratio (SMSE). We define $\text{SMSE}^{V}_{(M,U)}$ as the modeling mismatch between the transmitted signal $y(t)$ and $y^{V}_{(M,U)}(t)$ :
\begin{equation*}
\text{SMSE}^{V}_{(M,U)} = \frac{|y(t)|^2}{|y(t)-y^{V}_{(M,U)}(t)|^2}.
\end{equation*}
Similarly, we define $\text{SMSE}^{O}_{(M,U)}$ as the modeling mismatch between the signal in \eqref{digipower} and $y^{O}_{(M,U)}(t)$:
\begin{equation*}
\text{SMSE}^{O}_{(M,U)} = \frac{|y(t)|^2}{|y(t)-y^{O}_{(M,U)}(t)|^2}.
\end{equation*}

To calculate the $\text{SMSE}^{V}_{(M,U)}$ and $\text{SMSE}^{O}_{(M,U)}$ we simulate the optical communication system shown in Fig. \ref{fig:phy_sys} and calculate the $2^{nd}$-order Volterra Kernels and the orthogonal Volterra Kernels with different values of $\ell$ and $u$. In this simulation, OOK symbols ($a_k \in {0,1}$) are shaped using an unchirped Gaussian envelope $e^{-{t^{2}}/{2T_0^2}}$ with $T_0=36$ ps. The optical channel is modeled as \eqref{eq:fiber}, where $\lambda=1550$ nm is the wavelength, $DL = 600 $ ps/nm is the fiber dispersion and $C=3 \cdot 10^8$ m/s is the speed of light. Table \ref{tab:2ndVK} shows the values for $\text{SMSE}^{V}_{(M,U)}$ and Table \ref{tab:OVK} shows the values for $\text{SMSE}^{O}_{(M,U)}$.

\begin{table}[!t] 
\renewcommand{\arraystretch}{1.3}
\caption{Signal to MSE ratio for the $2^{nd}$ order Volterra Kernel model}
\begin{center}

\begin{tabular}{ c | c c c }\label{tab:2ndVK}

$\text{SMSE}^{V}_{(M,U)} (dB)$ & $M=3$ & $M=4$ & $M=5$\\
\hline
 U = 1 & 2.6191 & 2.6203 & 2.6092\\
 U = 2 & 7.7021 & 7.7410 & 7.7229\\
 U = 3 & 16.5685 & 16.4693 & 16.3455\\
 U = 4 & -      & 27.0083 & 27.0452\\
 U = 5 & -      & -       & 46.1393\\
\end{tabular}
\end{center}
\end{table}

\begin{table}[!t] 
\renewcommand{\arraystretch}{1.3}
\caption{Signal to MSE ratio for Orthogonal Volterra Kernel}
\begin{center}

\begin{tabular}{ c | c c c } \label{tab:OVK}

$\text{SMSE}^{O}_{(M,U)} (dB)$ & $M=3$ & $M=4$ & $M=5$ \\
\hline
 U = 1 & 15.0743 & 19.5336 & 30.9902 \\
 U = 2 & 16.2693 & 24.2276 & 37.3002\\
 U = 3 & 16.5685 & 26.9436 & 42.2060\\
 U = 4 & -       & 27.0083 & 45.7625\\
 U = 5 & -       & -       & 46.1393\\
\end{tabular}
\end{center}
\end{table}

Comparing Tables \ref{tab:2ndVK} and \ref{tab:OVK}, we see that SMSE values increase as the number of kernels considered in the expansion increases; this happens for both models and shows the importance of selecting an $M$ sufficiently large so the modeling mismatch does not become a dominant source of noise to the MLSE-based receiver. As expected, for fixed $U$ the values are essentially independent of $M$. In the case of the orthogonal Volterra Kernel model, one kernel in the receiver is sufficient to achieve satisfactory BER performance (see below).

The fact that the orthogonal Volterra Kernel model concentrates most of the features of the transmitted signal allows the MLSE-based receiver to function well with implementing fewer branches in the filter-bank structure. This is very important in terms of complexity for the receiver, specially for short-range optical communication applications where the number of multiplications required by the \ac{VD} is of the same order of magnitude required by multiple linear filtering operations performed by the filter bank.

\section{MLSE-Based Receivers Design}

In \cite{maggio2014reduced}, the branch metric implemented in the decoder reflects the orthogonal Volterra Kernel channel model and the receiver structure. It takes into consideration the kernels and the matched and whitening filters to calculate an equivalent channel that in turn is used to evaluate the Euclidean distance between the received symbol and the calculated sequences. We define $c_u$ as the equivalent response of branch $u=1,2,...,U$ at the input of the \ac{VD} at symbol time:
\begin{equation*}
c_u[i]=h_u[i]\ast h_u[-i]*w_u[i],
\end{equation*}
where $h_u[-i]$ and $w_u[i]$ are the matched and whitening filters of branch $u$. The branch metric proposed in \cite{maggio2014reduced} is defined as a multidimensional Euclidean distance:
\begin{equation*}
\sigma=|| \mathbf{r}[k]-\mathbf{c}[k]\ast \mathbf{b}_k||^2,
\end{equation*}
where $\mathbf{r}[k]=[r_1[k],...,r_u[k]]^T$ is composed of the received symbols at each branch, $\mathbf{c}[k]=[c_1[k],...,c_u[k]]^T$ is composed of the equivalent responses of each branch and $\mathbf{b}_k=[b_k^{(0)},...,b_k^{(u)}]^T$.

In applications where computational complexity is of great importance, the implementation of an optimal receiver as depicted in Fig. \ref{fig:st_wmf_mlsd} becomes unfeasible, and the necessity of truncating the system model when designing the receiver arises. For example, to exactly match the system model expanded in $M=U$ kernels, and assuming that matched and whitening filters have the same length at different branches, we can calculate the number of multiplications required for each detected symbol: 
\begin{equation*}
U(L_{MF}+L_{WF}+A^{L_{VD}}),
\end{equation*} 
where $L_{MF}$ denotes the length of the matched filters at symbol time, $L_{WF}$ denotes the length of the whitening filters, $A$ is the constellation size and $L_{VD}$ denotes the memory of the \ac{VD}. Note that the number of multiplications grows linearly with the number of branches considered in the truncated model. In this sense, given the results in Section \ref{sec:parameters} and in interest of complexity, it is reasonable to design the MLSE-based receiver with only one branch, as long as the number of expanded kernels is satisfactory. We call this MLSE-based receiver structure: Volterra Pre-Filtering + $\sigma$-metric (VPF+$\sigma$). 

However, in such scheme, even when the modeling mismatch is not very significant, the BER performance might suffer, given that the model is also taken into consideration at each branch metric calculation. As result, a small error can be propagated through the trellis. Another alternative is to use a different metric at the \ac{VD} that does not take into consideration a specific model of the optical channel. In this paper, we also consider a metric based on the average value of the possible received sequences, which we call the $\mu$-metric:
\begin{equation*}
\mu = \frac{\sum_k^{K} d_k}{\sum_k^{K} \mathcal{I} \{ \mathbf{a}_k = \{a_k,..., a_{k-L_{VD}+1} \} \}}
\end{equation*}
\begin{equation*}
d_k = \begin{cases} r_k & \mbox{if } \mathcal{I} \{ \mathbf{a}_k = \{a_k,..., a_{k-L_{VD}+1}\} \} = 1 \\ 
0 & \mbox{if } \mathcal{I} \{ \mathbf{a}_k = \{a_k,..., a_{k-L_{VD}+1}\} \} =0 \end{cases}
\end{equation*}
where $r_k$ is the received symbol at instant $k$, $K$ is the total number of training symbols, and $\mathcal{I}$ indicates when  the sequence $\{a_k,..., a_{k-L_{VD}+1} \}$ is present at the training sequence.

Finally, taking advantage of the Orthogonal Volterra Kernel model of the optical channel and trading off complexity and modeling mismatch, we propose a third MLSE-based receiver structure: the Volterra Pre-Filtering + $\mu$-metric (VPF+$\mu$). In this scheme, we use the information of the Orthogonal Volterra Kernel model to design the matched and whitening filter pre-filtering scheme and implement the $\mu$ metric at the \ac{VD} to decouple the modeling error of the decoding process.

\section{Numerical Simulations}

\begin{figure}[t!]
  \centering
    \includegraphics[width=0.5\textwidth]{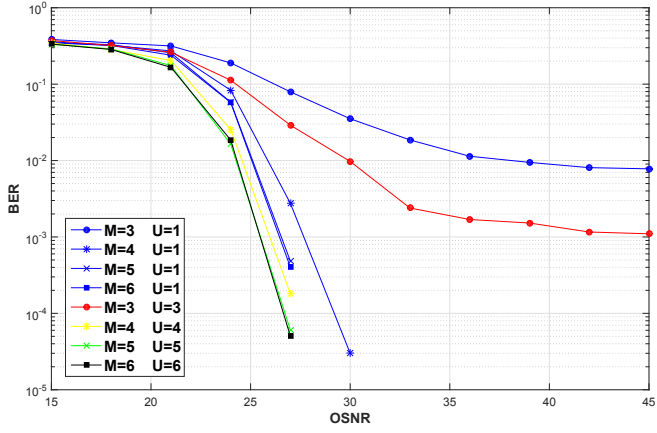}
 \caption{BER performance analysis obtained for different number of orthogonalized kernels, $M=3,...,6$, and different number of truncated kernels, $U=3,...,6$, using OOK modulation. The blue curves show the performance of receivers implemented with only one branch at the input of the \ac{VD} detector. The other curves show the performance of receivers implemented with the maximum possible number of branches at the input of the detector.}
\label{fig:kernels_used}
\end{figure}

To investigate the impact on performance of the usage of different number of branches in the receiver structure, we simulated the optical communication system with the same parameters as the simulation scenario that generated Tables \ref{tab:2ndVK} and \ref{tab:OVK} and implemented the receiver shown in Fig. \ref{fig:st_wmf_mlsd} with a variable number of branches in the filter bank. Fig. \ref{fig:kernels_used} shows the BER performance achieved by MLSE-based receivers implemented based on models expanded to $3$, $4$, $5$ and $6$ kernels using different number of branches in the filter bank at the input of a \ac{VD} decoder. In the plot, the number of kernels used in the expansion model are represented in the legend by $M$, while the number of branches used in the filter bank is represented by $U$. The \ac{VD} was implemented with a memory of $5$.

\begin{figure}[h!]
  \centering
    \includegraphics[width=0.5\textwidth]{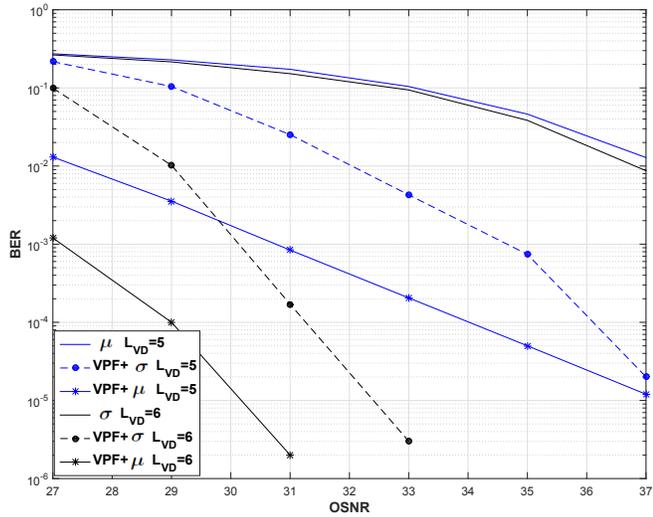}
 \caption{BER performance for OOK modulation implementing the three design options consided: VPF+$\sigma$, VPF+$\mu$ and the MLSE-based receiver with no pre-filtering scheme and the $\mu$ metric implemented at the \ac{VD}. The curves in blue were obtained using a \ac{VD} with memory of $5$ while the curves in black were obtained with a \ac{VD} with memory of $6$. }
\label{fig:OOK_comp}
\end{figure}

The BER curves in Fig. \ref{fig:kernels_used} reaffirm the importance of choosing a high enough number of kernels to model the transmitting signal, since receivers implemented based on models expanded on $3$ kernels were significantly outperformed by receivers with models expanded in $4$, $5$ and $6$ kernels. On the other hand, it is possible to say that the number of kernels used in the receiver filter-bank does not have a significant impact in the performance, given that the BER curves are shown to have similar performances even for the case where only $1$ branch was used in the receiver for $5$ and $6$ kernels expansion models. This result motivates the design of less complex receivers by discarding unnecessary filtering devices.

In order do compare the different MLSE-based designs presented in this paper, we simulate the optical communication system according to Fig. \ref{fig:phy_sys}. In this case, we excite the channel with both OOK and 4-PAM symbols and simulate the fiber with dispersion of $DL = 800 $ ps/nm and symbol rate $1/T=28GHz$. The other parameters are the same as the previous scenario. Next, we present the simulated BER performances of the three considered receiver designs.

Fig. \ref{fig:OOK_comp} shows that the receivers without any pre-filtering scheme were significantly outperformed by those with matched and whitening filters calculated from the Orthogonal Volterra Kernel model. The MLSE-based receivers with pre-filtering scheme and $\mu$ metric outperformed the receivers with $\sigma$ metric by approximately 4 dB with a \ac{VD} of memory $5$ and by 3 dB with a \ac{VD} of memory $6$ at a BER of $10^{-3}$. This shows that modeling mismatch caused by using only $1$ branch in the pre-filtering scheme can be compensated by decoupling the branch metric of the model, improving the final BER performance.

\begin{figure}[t!]
  \psfrag{m}{$\mu$}
   \includegraphics[width=0.5\textwidth]{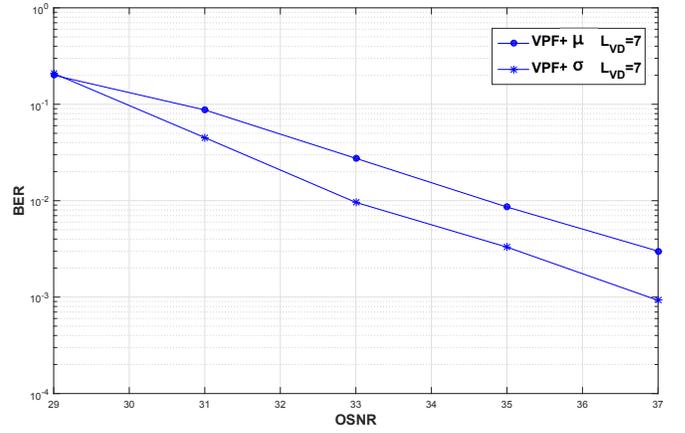}
 \caption{BER performance for 4-PAM modulation. The curves represent the performances of MLSE-based receivers implemented with a $6$ kernels expansion and with \ac{VD}'s with memory $7$.}
\label{fig:PAM4_comp}
\end{figure}

Finally, in Fig. \ref{fig:PAM4_comp}, we see that for 4-PAM, the VPF+$\mu$ scheme once again outperforms VPF+$\sigma$ scheme by approximately 2 dB at a BER of $10^{-2}$. This reaffirms the superiority of the proposed VPF+$\mu$ in comparison to the two other considered schemes in terms of both performance and complexity. In addition, this result shows the usability of such MLSE-based receivers in higher data rate scenarios with reasonable complexity, moving closer to the high capacity and low cost goals of the optical communications industry.

\section{Conclusion}
In this paper, we considered MLSE-based receivers using two different Volterra expansion models. We have shown that the model based on orthogonal kernels allows for a truncation to a smaller number of terms for given modeling mismatch compared to the non-orthogonal model. A single prefiltering branch turns out to be sufficient in practice, lowering the complexity at the receiver.
We also provided three MLSE-based receiver design options: a receiver without pre-filtering, one with Volterra pre-filtering and $\sigma$-metric (VPF+$\sigma$) and one with Volterra Pre-Filtering + $\mu$-metric (VPF+$\mu$). BER performance results show the superiority of the proposed VPF+$\mu$ in comparison to the two other considered schemes in terms of both performance and complexity. These results apply to both OOK and 4-PAM modulation schemes.

\bibliography{IEEEabrv,omlsebib}
\bibliographystyle{IEEEtran}

\end{document}